\documentclass[aps,prl,showpacs,twocolumn,superscriptaddress,10pt]{revtex4-1}

\usepackage{amsmath}
\usepackage{graphicx}
\usepackage{bm}

\newcommand{\eloc}{E_\mathrm{loc}}
\newcommand{\dloc}{\delta_\mathrm{loc}}

\newcommand{\loc}{\mathrm{loc}}

\usepackage{color}

\date{\today}

\begin{document}

\title{Chaos-assisted tunneling in the presence of Anderson localization}
\author{Elmer V. H. Doggen}
\email[Corresponding author: ]{doggen@irsamc.ups-tlse.fr}
\affiliation{Laboratoire de Physique Th\'eorique, IRSAMC, Universit\'e de Toulouse, CNRS, UPS, France}
\author{Bertrand Georgeot}
\affiliation{Laboratoire de Physique Th\'eorique, IRSAMC, Universit\'e de Toulouse, CNRS, UPS, France}
\author{Gabriel Lemari\'e}
\affiliation{Laboratoire de Physique Th\'eorique, IRSAMC, Universit\'e de Toulouse, CNRS, UPS, France}
\affiliation{Department of Physics, Sapienza University of Rome, P.le A. Moro 2, 00185 Rome, Italy}

\begin{abstract}
Tunneling between two classically disconnected regular regions can be strongly affected by the presence of a chaotic sea in between.
This phenomenon, known as chaos-assisted tunneling, gives rise to large fluctuations of the tunneling rate.
Here we study chaos-assisted tunneling in the presence of Anderson localization effects in the chaotic sea. 
Our results show that the standard tunneling rate distribution is strongly modified by localization, going from the Cauchy distribution in the ergodic regime to a log-normal distribution in the strongly localized case, for both a deterministic and a disordered model.
We develop a single-parameter scaling description which accurately describes the numerical data.
Several possible experimental implementations using cold atoms, photonic lattices or microwave billiards are discussed.
\end{abstract}

\maketitle

Tunneling has been known since early quantum mechanics as a striking example of a purely quantum effect that is classically forbidden.
However, the simplest presentation based on tunneling through a one-dimensional barrier does not readily apply
to higher dimensions or to time-dependent systems, where the dynamics becomes more complex with various degrees of chaos \cite{bohigas1993manifestations}. 
An especially spectacular effect of chaos in this context is known as chaos-assisted tunneling \cite{tomsovic1994chaos}: in this case, tunneling is mediated by ergodic states in a chaotic sea, and tunneling amplitudes have reproducible fluctuations by orders of magnitude over small changes of a parameter.
This is reminiscent of universal conductance fluctuations which arise in condensed matter disordered systems \cite{akkermans2007mesoscopic}.

A typical example of chaos-assisted tunneling arises for systems having a mixed classical dynamics where regular zones with stable trajectories coexist with chaotic regions with ergodic trajectories. 
In the presence of a discrete symmetry, one can have two symmetric regular zones classically disconnected, i.e.\ classical transport between these two zones is forbidden. 
However, quantum transport between these two structures is possible through what is called dynamical tunneling \cite{davis1981quantum}. 
Regular eigenstates are regrouped in pairs of symmetric and antisymmetric states whose eigenenergies differ by a splitting $\delta$ inversely proportional to the characteristic tunneling time.  
If a chaotic region is present between the two regular structures, tunneling becomes strongly dependent on the specific features of the energy and phase-space distributions of the chaotic states. 
It results in strong fluctuations of the tunneling splittings which are known to be well-described by a Cauchy distribution \cite{Leyvraz1996a}. 
This has been extensively studied, both theoretically \cite{aaberg1999chaos, mouchet2001chaos, mouchet2003signatures, artuso2003effects, wuster2012macroscopic, Backer2010a, keshavamurthy2011dynamical} and experimentally \cite{dembowski2000first, hofferbert2005experimental, backer2008dynamical, dietz2014spectral, gehler2015experimental, steck2001observation, hensinger2001dynamical, hensinger2004analysis, lenz2013dynamical}.

However, states in a chaotic sea are not necessarily ergodic; indeed, in condensed matter, quantum interference effects can stop classical diffusion and induce Anderson localization \cite{Anderson1958a, evers2008anderson, abrahams201050}. 
In chaotic systems, a similar effect known as dynamical localization occurs where states are exponentially localized with a characteristic localization length $\xi$ \cite{casati1979stochastic, grempel1984quantum, Moore1995a, Lemarie2008a}. 
When the size $L$ of the chaotic sea is much smaller than $\xi$, chaotic states are effectively ergodic, whereas a new regime arises when $\xi \ll L$ where strong localization effects should change the standard picture of chaos-assisted tunneling. 
Indeed, it is well-known for disordered systems that in quasi-1D universal Gaussian conductance fluctuations are replaced by much larger fluctuations in the localized regime with a specific characteristic log-normal distribution \cite{Beenakker1997a}.
Moreover, for chaotic systems, it was shown in \cite{Ishikawa2009a} that tunneling is drastically suppressed by dynamical localization, and that the Anderson transition between localization and diffusive transport manifests itself as a sharp enhancement of the average tunneling rate \cite{ishikawa2010dynamical}.

In this paper, we extend the theory of chaos-assisted tunneling to the previously unexplored regime where Anderson localization is present in the chaotic
sea. 
We show that the distribution of level splittings changes from the Cauchy distribution in the ergodic regime to a log-normal distribution in the strongly localized regime.
We consider two models to study these effects: a deterministic model having a mixed phase space and whose quantum chaotic states display dynamical localization, and a disordered model based on the famous Anderson model \cite{Anderson1958a}.
This allows us to study numerically the two extreme ergodic and localized regimes as well as the crossover between them. 
The numerical data are found to follow a one-parameter scaling law with $\xi/L$. 
We present simple analytical arguments which account for the observed behaviors throughout the full range of parameters for both models.
This shows that in the presence of Anderson localization the fluctuations of chaos-assisted tunneling show a new universal behavior.
Our theory is generic in 1D, and the approach should be generalizable to higher dimensions.

\begin{figure*}[!htb]
 \includegraphics[width=0.8\linewidth]{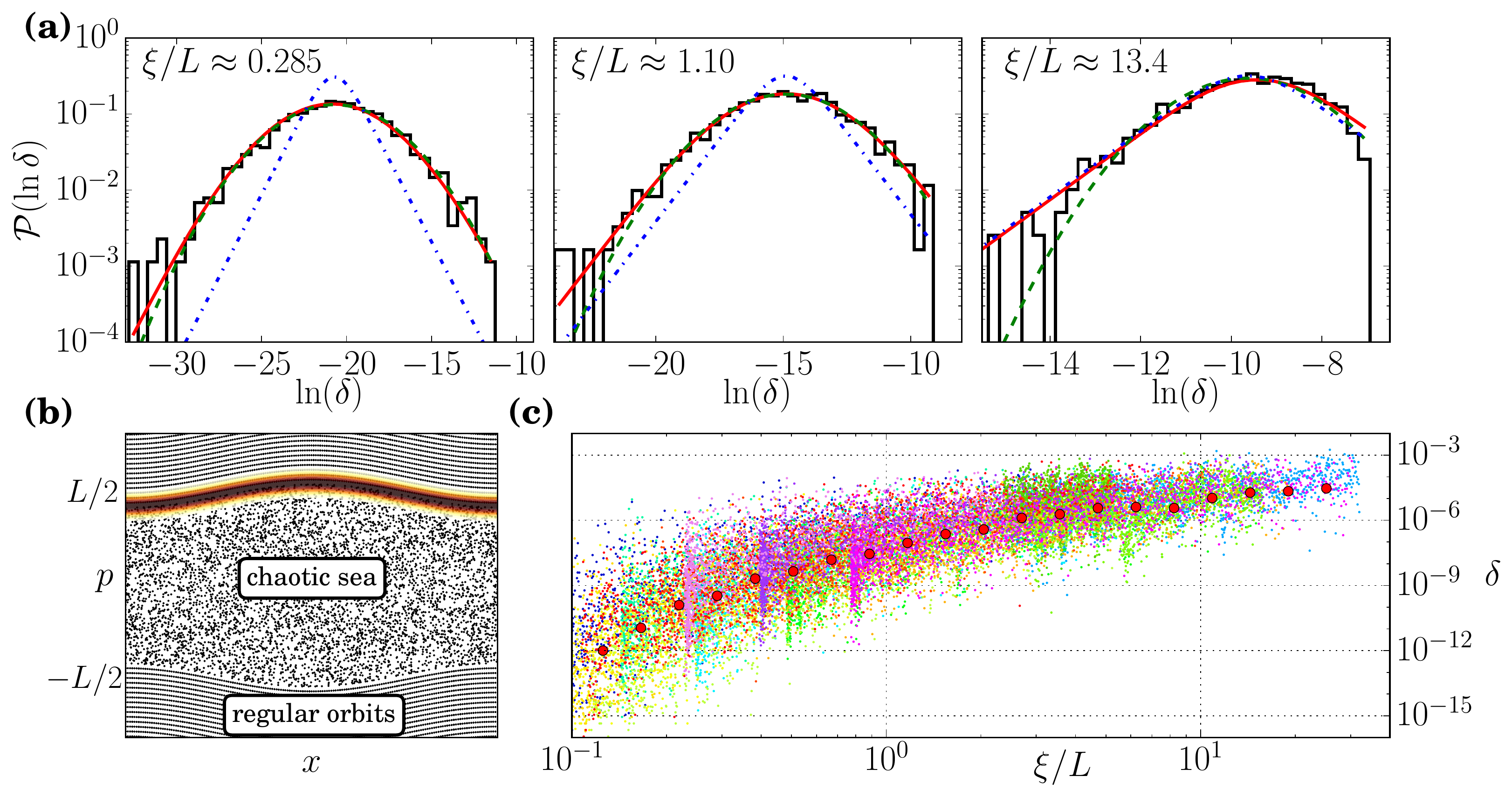}
 \caption{(Color online). \textbf{(a)} Splitting distribution $\mathcal{P}(\ln \delta)$ for the deterministic model \eqref{eq:hamiltonian} in the localized (left), intermediate (middle) and ergodic (right) regimes. 
 Green dashed line: fit with a log-normal distribution.
 Blue dashed-dotted line: fit by the Cauchy distribution. 
 Red solid line: analytical theory Eq.\ \eqref{eq:combined_distribution} with $\Gamma$ and $\delta_\text{typ}$ fitting parameters (see Fig.~\ref{fig:anderson}b).
 In the localized (ergodic) regime the log-normal (Cauchy) distributions overlap almost perfectly with Eq.\ \eqref{eq:combined_distribution}.
 The data correspond to $K=30$, $p_\mathrm{r} = 0.4n$ ($n=2048$) and $6400$ values of $\hbar$ in a small range around $\hbar = 0.85, 0.45, 0.25$ respectively.
 \textbf{(b)} 
  Phase-space (Husimi) distribution of the initial wave function (color shades) superimposed on the classical phase space of the deterministic model \eqref{eq:hamiltonian}, with $\hbar = 0.25$, $K = 30$, $p_\text{c} = 10$ and $p_\mathrm{r} = 1024$. 
 \textbf{(c)} Splittings $\delta$ as a function of $\xi/L$ showing single-parameter scaling behavior. 
 There are 16 sets (different colors) corresponding to fixed values of $K = 20, 30, 40, 50$ and $p_\mathrm{r}$ varies from $0.1n$ to $0.7n$ ($n=2048$ is the system size) for 1600 values of $\hbar \in [0.1,1.5]$.
 Big red dots: typical value of $\delta$ averaged over all data sets.}
 \label{fig:scalingplot}
\end{figure*}

\emph{Models.---}
In order to observe standard chaos-assisted tunneling, the system should have two regular states 
(symmetric and anti-symmetric) weakly coupled via a large set of other states, whose statistics is well described by Random Matrix Theory.
In our case, we need these states to be Anderson localized (e.g.\ well described by random band matrices \cite{Casati1990b, Fyodorov1991}) in the direction relevant to the discrete symmetry of the system.
 We will focus on two specific 1D models.

The deterministic model that we use is a variant introduced in \cite{Ishikawa2009a} of the quantum kicked rotor, a paradigmatic model of quantum chaos which displays
dynamical localization in momentum space.
The Hamiltonian is given by:
\begin{equation}
 \hat{H} = T(p) + K \cos(x) \sum_{n=-\infty}^\infty \delta(t-n), \label{eq:hamiltonian}
\end{equation}
where $p \hbar$ is momentum, $x$ is a dimensionless position (or phase) with period $2 \pi$, $t$ is time and $K$ is the kick strength.
The dispersion relation $T(p)$ is such that the phase space exhibits well-separated chaotic and regular regions \cite{Ishikawa2009a}:
\begin{subequations}
\label{eq:disp}
\begin{align}
 T(p) = \,& \frac{(\hbar p)^2}{2} & \text{for} \, |p| \leq p_\mathrm{r}/2, \label{eq:disp1}\\
 T(p) = \,& \omega |p| + \omega_0& \text{for} \, |p| > p_\mathrm{r}/2. \label{eq:disp2}
\end{align}
\end{subequations}
The value of $\omega$ should be irrational, throughout this work we use $\omega = 2\sqrt{5} \approx 4.4721\dotso$.
$\omega_0$ is chosen such that $T(p)$ is continuous.
Although the dispersion relation may look artificial, we will show that a similar system could be implemented in a realistic setting.
A discrete symmetry $p \rightarrow -p$ exists.
The region of phase space $|p| \leq p_\mathrm{r}/2$ is strongly chaotic for $K\gg1$. 
The characteristic size of the chaotic sea is $L = p_\mathrm{r} + K/[\hbar \sin(\omega/2)]$ (see Fig.\ \ref{fig:scalingplot}b).
The two regions $|p| > L/2$ correspond to two momentum symmetric regular zones.  
In the following, we consider an initial state located on a classical torus in the regular part of phase space as obtained using Einstein-Brillouin-Keller quantization (see \cite{Ishikawa2009a}), with initial mean momentum $\langle p \rangle (t=0)= p_\text{c}+ L/2$. 
The distance to the chaotic sea $p_c$ is set to $p_c = 10$ such that the coupling to the chaotic sea is constant when varying $\hbar$.
In Fig.\ \ref{fig:scalingplot}b we portray this initial state superimposed on the classical phase space of \eqref{eq:disp}.

\begin{figure*}[!htb]
 \includegraphics[width=1.0\linewidth]{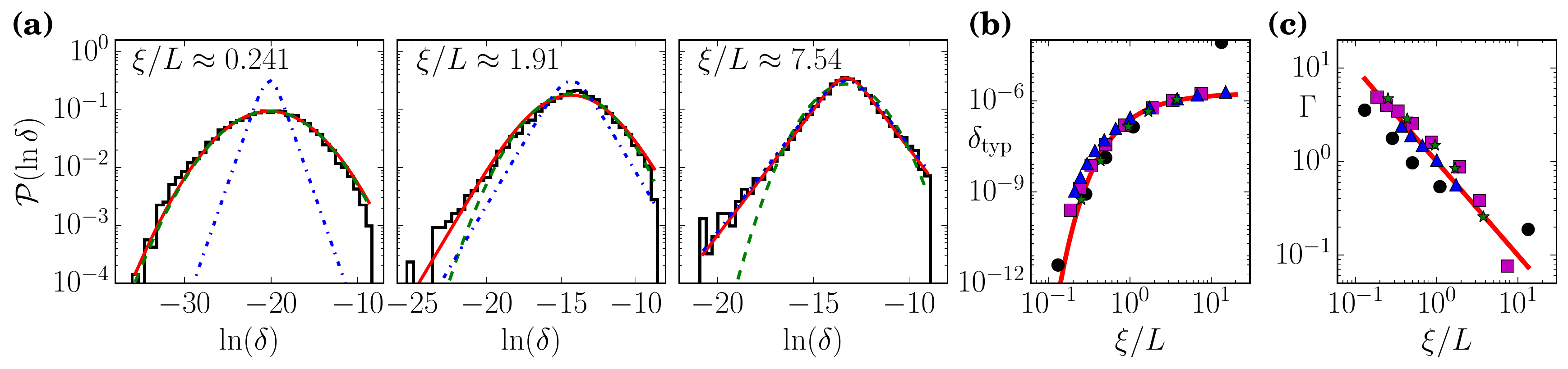}
 \caption{
 (Color online). \textbf{(a)} Splitting distribution $\mathcal{P}(\ln \delta)$ for the disordered model \eqref{eq:anderson}, for $W = 3, 1, 1/2$ from left to right.
 Note the strong similarity with the distributions for the deterministic model \eqref{eq:hamiltonian} of Fig.\ \ref{fig:scalingplot}a.
 Colored lines: as in Fig. \ref{fig:scalingplot}.
 $L=256$, $t_c = 0.001$, $b = 3$, and $10000$ realizations (see text).
 \textbf{(b)} 
 Scaling behavior of $\delta_\text{typ}$ as a function of $\xi/L$.
 Disordered model \eqref{eq:anderson}: $L = 128$ (blue triangles), $L =256$ (magenta squares) and $L = 512$ (green stars), parameters as in \textbf{(a)}.
 Black circles: deterministic model \eqref{eq:hamiltonian}, $K=30$, 6400 values of $\hbar$ in a small range around $0.25,0.45,0.65,0.85,1.45$, $p_\mathrm{r} = 0.4n$ $(n=2048)$.
 Red solid line: theoretical prediction $\delta_\mathrm{typ} = \delta_\mathrm{ch,typ} \exp(-2L/\xi)$, with $\delta_\mathrm{ch,typ}$ fitted to $L=512$ data.
 \textbf{(c)} Scaling behavior of $\Gamma$ (see Eq.\ \eqref{eq:combined_distribution}) as a function of $\xi/L$ (symbols defined as in the left panel). 
 Red solid line: theoretical prediction $\Gamma = L/\xi$.
 }
 \label{fig:anderson}
\end{figure*}

The second model we use is a variant of the Random Matrix model of chaos-assisted tunneling \cite{Leyvraz1996a} in a condensed matter context, allowing for localization effects.
It is a disordered system based on the Anderson model:
\begin{align}
 \hat{H}_\text{A} & =  \sum_{i \neq 1,L} w_i a^\dagger_i a_i + \sum_{\langle i,j \rangle} a^\dagger_i a_j + \text{H.c.} \nonumber \\ 
&+ t_c (a^\dagger_1 a_2 + a^\dagger_L a_{L-1} + \text{H.c.}) \label{eq:anderson}
\end{align}
where the summation $\langle i,j \rangle$ is over indices $1 \leq |i-j| \leq b$ ($i,j \neq 1,L$), $b \geq 1$ an integer, $L$ is the lattice size and $a^{(\dagger)}$ are annihilation (creation) operators.
The on-site energies $w_i$ are independent random numbers, uniformly distributed $\in [-W/2,W/2]$, with the constraint that the $w_i$'s are \emph{symmetric} with respect to the lattice center (analogous to the discrete symmetry $p\rightarrow -p$ for model \eqref{eq:hamiltonian}).
The coefficient $t_c \ll 1$ describes the coupling of regular states (sites $i = 1,L$) to the analog of the chaotic states (sites $i = 2\ldots L-1$).
Tunneling beyond nearest-neighbor with $b>1$ is required to reach a regime where the statistics of the analog of the chaotic states is close to that of GOE Random Matrices \cite{Casati1990b, Fyodorov1991, footnoteBRM} (the standard Anderson model permits only localized and ballistic behavior). 
In the following we use $b=3$, but we have checked that other values lead to similar results (data not shown).

In these two models, there exists a characteristic localization length  $\xi=\langle 1/\xi_\mathrm{loc}\rangle^{-1}$ defined through the average over individual localization lengths $\xi_\mathrm{loc}$ of the chaotic states \cite{Beenakker1997a,BoltonHeaton1999a}. 
In the model \eqref{eq:hamiltonian}, the Floquet eigenstates in the chaotic sea are exponentially localized in momentum space, with $\xi \approx (K^2/4\hbar^2)[1 - 2J_2(\tilde{K})(1 - J_2(\tilde{K}))]$ ($J_2$ denotes the Bessel function and $\tilde{K} = (2K/\hbar)\sin \hbar/2$) \cite{Shepelyansky1987a}. 
 In the model \eqref{eq:anderson}, it is known that for $b = 1$ the eigenstates are exponentially localized in position space, with a localization length near the band center $\xi_\mathrm{1D}^{-1} \approx \ln (1 + W^2/16)/2 + 4 \arctan(W/4)/W - 1$ \cite{Izrailev1998a}.
 For $1<b\ll L$ the system is still localized, with numerically $\xi \approx (2b - 1) \xi_\mathrm{1D}$.

\emph{Splitting statistics for the deterministic model.---}
The dynamics of \eqref{eq:hamiltonian} can be integrated over one period to give the evolution operator $
 \hat{U}~=~e^{-\frac{i}{2\hbar} T(p)} e^{-\frac{i}{\hbar} K \cos(x)} e^{-\frac{i}{2\hbar} T(p)}$.
 Tunneling may be studied through the Floquet eigenstates $\vert \Psi_\alpha\rangle $ of $\hat{U}$ with quasi-energies $\varepsilon_\alpha$ obeying $\hat{U} \vert \Psi_\alpha\rangle = e^{i \varepsilon_\alpha} \vert \Psi_\alpha\rangle $.
Classically, transport to the chaotic sea or the other regular island is forbidden by the presence of invariant curves.
However, in the quantum regime, the initial state, having a certain expectation value of momentum $\langle p \rangle (t=0)= p_\text{c}+ L/2$, will tunnel through the chaotic sea to the other side with an oscillation period $T_\text{osc}$, so that $\langle p \rangle (T_\text{osc}/2) = -\langle p \rangle(t=0)$.
$T_\text{osc}$ can be identified with $2\pi/\delta$, where $\delta=\varepsilon^+ - \varepsilon^-$ is the splitting between the symmetric and anti-symmetric Floquet eigenstates having the largest overlap with the initial state.

In Fig.\ \ref{fig:scalingplot}a we show the distributions of $\ln \delta$ in the different regimes of the chaotic sea: localized $\xi\ll L$, ergodic $\xi\gg L$ and intermediate $\xi \approx L$.
One can see that the distributions are markedly different: we recover the known Cauchy distribution in the ergodic regime whereas a log-normal distribution is observed in the localized regime. In the crossover regime, an intermediate distribution is obtained with a non-trivial shape to be discussed later. 
In order to reveal which scales control these behaviors, we considered the splittings $\delta$ for many different parameters (see caption of Fig.\ \ref{fig:scalingplot}).
In Fig.~\ref{fig:scalingplot}c we show that, strikingly, the data follow a single-parameter scaling law as a function of $\xi/L$.
The essential physics of the problem therefore only depends on the localization length $\xi$.

\emph{Splitting statistics for the disordered model. ---} 
The disordered model \eqref{eq:anderson} has no obvious classical analog.
Nevertheless, it can be used to describe chaos-assisted tunneling, with the localization length in position space, dependent on $W$, analogous to the localization length in momentum space of model \eqref{eq:hamiltonian}, dependent on $K$ and $\hbar$.
The splitting $\delta$ is determined by computing the eigenfunctions most strongly overlapping with the first site (due to symmetry, this is equivalent to the last site).
The splitting distributions for this model are shown in Fig.\ \ref{fig:anderson}a. 
Strikingly, very similar behavior is observed in this disordered model compared to model \eqref{eq:hamiltonian}.
We recover the Cauchy distribution at small values of $W$ where the disordered states are delocalized, whereas for large $W$ (localized states) the distribution has a log-normal shape, with again an intermediate behavior at the crossover.
In Fig.\ \ref{fig:anderson}b we show, for both the deterministic and disordered models, the scaling behaviors of $\delta_\mathrm{typ} = \exp\langle\ln \delta\rangle$, and in Fig.\ \ref{fig:anderson}c the fitted parameter $\Gamma$ related to the width of the distribution (see Eq.\ \eqref{eq:combined_distribution}).

\emph{Analytical arguments.---}
In this section, we derive an expression for the splitting distribution valid in all regimes, provided the system displays the essential characteristics explained above.
Following \cite{Leyvraz1996a}, the splitting $\delta$ between the symmetric and antisymmetric regular states can be obtained from the displacements $\delta^{\pm}$ of their energies due to the coupling to the chaotic/disordered sea. 
Because this coupling $v$ is classically forbidden, it is exponentially small and a perturbation theory leads to: $\delta^{\pm}\approx\sum_i \delta^{\pm}_i$ with 
$ \delta^{\pm}_i={v_i}^2/(E-E_i^{\pm})$, $E_i^{\pm}$ being the energy of the chaotic state indexed by $i$ of the same symmetry as the regular state of energy $E$ \footnote{Because the regular orbits are strongly localized far apart from each other, we can neglect their direct splitting and assume that $E=E_\mathrm{reg}^+ = E_\mathrm{reg}^-$.}.
The splitting $\delta$ is then $\delta=\vert \delta^{+}-\delta^-\vert$. 
In the delocalized ergodic regime, the overlap between chaotic states is large which excludes that symmetric states are resonant with antisymmetric states, thus $\delta^{+}$ and $\delta^{-}$ are uncorrelated. 
The splitting distribution is then given by the distribution of $\delta^{\pm}$, and follows a Cauchy law \cite{Leyvraz1996a}.
In the following, we will focus on the regime of large splittings obtained when 
a single resonant term  of energy $E_\mathrm{ch}^{\pm}$ dominates the sum: $ \delta^{\pm}\approx v^2/ (E - E_\mathrm{ch}^{\pm})$.

Localization of the wave functions in the
chaotic sea comes into play in two ways : it can i) induce a strong correlation between the
energies of the symmetric $( ^+ )$ and antisymmetric $( ^- )$ chaotic states, or ii) modify the
tunneling matrix element $v$. We now proceed by analyzing the distribution of $\delta$ in the two limiting cases where one mechanism is dominant.

When case i) is dominant, it is the strong correlation of $\delta^\pm$ which is at the origin of the departure of the splitting distribution from the Cauchy law (see \cite{tomsovic1994chaos, Leyvraz1996a} for similar correlations arising due to partial classical barriers in the chaotic sea). 
We define the auxiliary quantities $E_\mathrm{ch}^{\pm} = \eloc \mp \dloc/2$ with $\dloc$ the splitting of the localized chaotic states, $\dloc \approx \Delta \exp(-2 L_\mathrm{loc}/\xi_\mathrm{loc})$ where $L_\mathrm{loc} $ is the distance between the two peaks of the localized (anti)symmetric chaotic state whereas $\Delta$ is the mean level spacing. 
We have $|\dloc| \ll |E - \eloc|$ in the correlated regime of interest, which yields: $\delta \approx \Big|\frac{v^2}{E-\eloc} \frac{\dloc}{E-\eloc}\Big|$.  
In the first factor, $v = v_\mathrm{ch} \exp(-2 L_\mathrm{reg}/\xi_\mathrm{loc})$ where $v_\mathrm{ch}$ is the usual coupling strength to the chaotic sea in the ergodic case, and
$\exp(-2 L_\mathrm{reg}/\xi_\mathrm{loc})$ describes the effect of localization of the chaotic state, located at a distance $L_\mathrm{reg}$ from the regular border. 
As $L=2 L_\mathrm{reg}+L_\mathrm{loc}$, the splitting can be approximated by:
\begin{equation}\label{eq:splitting}
 \delta \approx \delta_\mathrm{ch} \exp(-2 L/\xi_\mathrm{loc})  \; , 
\end{equation}
where we have taken $\vert E-\eloc\vert \sim \Delta$ which is the typical case in this regime \footnote{ {Note that strictly speaking, the splitting in the correlated case i) is given by $ \delta \approx {v_\mathrm{ch}}^2\Delta \exp(-2 L/\xi_\mathrm{loc})/(E-\eloc)^2 $. The $(E-\eloc)^2$ in the denominator may seem to change the tail of the distribution for rare small $E-\eloc$ values. However, the formula is valid for $\vert E-\eloc \vert \gg \delta_\mathrm{loc}$, and thus can describe only the central log-normal part of the distribution \eqref{eq:combined_distribution}. This regime corresponds therefore to the typical case where $\vert E-\eloc \vert \approx \Delta$.}} and $\delta_\mathrm{ch}=\vert {v_\mathrm{ch}}^2/(E-\eloc) \vert$ the ``standard'' chaos-assisted tunneling term.

On the other hand, when case ii) is dominant, the correlation between $\delta^\pm$ can be neglected and the distribution of $\delta$ is given by that of $ \delta^\pm$. Since $\delta_\mathrm{loc} \approx \Delta \exp(-2 L_\mathrm{loc}/\xi_\mathrm{loc})$ should be large in this uncorrelated limit, we must have $L_\mathrm{loc}\ll \xi_\mathrm{loc}$ and thus $L_\mathrm{reg} \approx L/2 \gg L_\mathrm{loc}$. The coupling to the regular borders is then exponentially weak, $v \approx v_\mathrm{ch} \exp(-2 L_\mathrm{reg}/\xi_\mathrm{loc})\approx v_\mathrm{ch} \exp(-L/\xi_\mathrm{loc})$. Then, using $\delta \approx v^2/ (E - E_\mathrm{loc})$, we remarkably obtain precisely the same Eq.\ \eqref{eq:splitting}.

Although Eq.~\eqref{eq:splitting} was obtained in certain limiting cases, it proves itself valid more generally. Indeed, let us assume that the first term of \eqref{eq:splitting} has the known Cauchy distribution $P_\text{ch}(\ln \delta_\text{ch}) = 2e^{\ln \hat{\delta}_\text{ch}}/[\pi(1+e^{2\ln \hat{\delta}_\text{ch}})]$, 
where $\hat{\delta} \equiv \delta/\delta_\text{typ}$.
In our case $\delta_\text{ch,typ} = \exp \langle\ln \delta_\mathrm{ch}\rangle$ depends only on $t_c$ or $p_c$ (see also \cite{Backer2010a}).
Because the inverse localization length $1/\xi_\mathrm{loc}$ in quasi-1D has a normal distribution with width $\propto 1/L$, the second term $\tilde{\delta}_\loc \equiv \exp(-2L/\xi_\mathrm{loc})$ of \eqref{eq:splitting}  
has a log-normal distribution, analogous to that of conductance in the strongly localized limit~\cite{Beenakker1997a,BoltonHeaton1999a}:
 $P_\loc(\ln \tilde{\delta}_\loc) ~=~\exp [-(\ln 1/\tilde{\delta}_\loc - 2\Gamma)^2/(8\Gamma)]  /(4 \Gamma  \sqrt{2\pi}) ,$
where $\Gamma = L/\xi$.
The total distribution $P(\ln \delta)$ can be obtained by convolution and is given by:
\begin{align}
 P(\ln \delta) = & \frac{1}{4} \exp\Big(-\ln \hat{\delta} + 2\Gamma\Big) \Big[ 1 + \mathrm{erf}\Big( \frac{-4\Gamma + \ln \hat{\delta}}{\sqrt{8\Gamma}} \Big) \nonumber \\
 & + \exp(2\ln \hat{\delta}) \mathrm{erfc} \Big( \frac{4\Gamma + \ln \hat{\delta}}{\sqrt{8\Gamma}} \Big)   \Big]. \label{eq:combined_distribution}
\end{align}
The expression \eqref{eq:combined_distribution} can be fitted to the distributions obtained numerically (see Fig.\ \ref{fig:scalingplot}a and Fig.\ \ref{fig:anderson}a). 
A very good agreement is found with the numerical data, in both the extreme localized and ergodic regimes where \eqref{eq:combined_distribution} describes a log-normal or Cauchy distribution, respectively. The intermediate regime ($\xi/L \approx 1$) is characterized by log-normal behavior in the center of the distribution, around $\delta = \delta_\text{typ}$, with Cauchy-type behavior in the tails. The fitted values of $\delta_\text{typ}$ and $\Gamma$ are represented in Fig.\ \ref{fig:anderson}b for both models. 
A good agreement is found with the expected behavior $\delta_\text{typ} = \delta_\mathrm{ch,typ} \exp(-2 L/\xi)$ and $\Gamma = L/\xi$ in the localized regime.
Note that we find that the high-$\delta$ cutoff of the splitting distribution is not affected by the localization and remains at the same value $\delta\approx v_\mathrm{ch}$ as in the ergodic case. This is clearly observed in Fig.\ \ref{fig:anderson} where the cutoff remains at $\delta\approx t_c$ in all regimes for the disordered model. This can be understood from \eqref{eq:splitting}: the cutoff is present in the distribution of $\delta_\mathrm{ch}$ while the localization effects described by the second term do not involve any cutoff in their distribution.

\textit{Experimental implementation.---}
The effects we have presented should be observable for a generic class of models sharing the properties listed above.
It is possible to implement a realistic version of the model \eqref{eq:hamiltonian} using a variant of the well-known cold-atom implementation of the quantum kicked rotor \cite{Moore1995a, Lemarie2008a, lemarie2009observation}. 
The dispersion relation \eqref{eq:disp} cannot be implemented directly. 
However, in the atomic kicked rotor, the kicking potential is usually realized through a sequence of short impulses of a stationary wave periodic in time. 
By truncating the Fourier series of this sequence at some specified harmonic $p_r$, one realizes a mixed system with a classically ergodic chaotic sea between $-p_r$ and $p_r$ \cite{blumel1986excitation, Chirikov1995} which realizes an experimentally accessible version of \eqref{eq:hamiltonian} (see \cite{dubertrand2016routes} for experimental details on an analogous system). 
The disordered model \eqref{eq:anderson} can be implemented readily using photonic lattices \cite{schwartz2007transport, lahini2008anderson, segev2013anderson}. 
A spatially symmetric disorder can be implemented in the direction transverse to propagation, and tunneling will result in oscillations in this transverse direction which can be easily measured. 
Lastly, both chaos-assisted tunneling \cite{dembowski2000first, hofferbert2005experimental, backer2008dynamical, dietz2014spectral, gehler2015experimental} and Anderson localization \cite{sirko2000observation, dembowski1999anderson} have been observed in microwave chaotic billiards. 
In this context, both disordered and deterministic models could be implemented and can yield very precise distributions of the tunneling rate.

\emph{Conclusion.---} We have shown that the presence of Anderson localization brings a new regime to chaos-assisted tunneling, with a new universal
distribution of tunneling rates. 
Remarkably, the crossover from the ergodic to the localized regimes is governed by a single-parameter scaling law, for both disordered and deterministic models.
Our theory accurately describes the different behaviors for our models, both of which could be implemented experimentally.
The results are generic in 1D in the localized regime, and the approach should be generalizable to higher dimensions.

Our study shows that chaos-assisted tunneling could be used as an experimental probe of the non-ergodic character of the chaotic sea. 
A fascinating perspective would be to generalize these ideas to other types of non-ergodic chaotic states, such as multifractal states, which are hard to see experimentally and have been intensively studied theoretically lately \cite{Mirlin1996a, evers2008anderson, Dubertrand2014}.

\begin{acknowledgments}
GL acknowledges an invited professorship at Sapienza University of Rome. We thank {\small{CALMIP}} for providing computational resources. This work was supported  by  Programme  Investissements  d'Avenir  under
the program ANR-11-IDEX-0002-02, reference ANR-10-LABX-0037-NEXT, by the ANR grant K-BEC No ANR-13-BS04-0001-01 and the ANR program BOLODISS.
Furthermore, we acknowledge the use of the following software tools and numerical packages: {\small{LAPACK}} \cite{Lapack1999}, the Fastest Fourier Transform in the West \cite{FFTW05} and Matplotlib \cite{Matplotlib}.
\end{acknowledgments}

\bibliography{ref}

\begin{thebibliography}{53}
\bibitem [{foo()}]{footnoteBRM}%
  \BibitemOpen
  \href@noop {} {}\bibinfo {note} {Strictly speaking, the limit of GOE
  statistics is reached in the limit $b\gg 1$ and $L\ll \xi$. Due to numerical
  restraints, we have considered values of b limited to $b=3,4,5$. This makes
  it possible to avoid the low disorder ballistic regime when $b = 1$ and $L\ll
  \xi$ and to recover the results expected for chaos-assisted tunneling in the
  ergodic regime.}\BibitemShut {Stop}%
\end{thebibliography}
\bibliographystyle{apsrev4-1}

\end{document}